\documentclass[11pt]{article}
\usepackage{epsfig}
\usepackage{amscd}
\usepackage{amssymb}
\usepackage{amsfonts}
\usepackage{amsmath}
\usepackage{graphics,epsfig,psfrag}

\begin{document}

\makeatletter
\@addtoreset{equation}{section}
\makeatother
\renewcommand{\theequation}{\thesection.\arabic{equation}}
\renewcommand{\thefootnote}{\alph{footnote}}

\begin{titlepage}

\baselineskip =15.5pt
\pagestyle{plain}
\setcounter{page}{0}

\begin{flushright}
\end{flushright}

\vfil

\begin{center}
{\huge On Koopman-von Neumann Waves II}
\end{center}

\vfil

\begin{center}
{\large E. Gozzi and D. Mauro}\\
\vspace {1mm}
Dipartimento di Fisica Teorica, Universit\`a di Trieste, \\
Strada Costiera 11, P.O.Box 586, Trieste, Italy \\ and INFN, Sezione 
di Trieste.\\
e-mail: {\it gozzi@ts.infn.it} and {\it mauro@ts.infn.it}
\vspace {1mm}
\vspace{3mm}
\end{center}

\vfil

\begin{abstract}
\noindent In this paper we continue the study, started in
\cite{waves1}, of the operatorial formulation 
of classical mechanics given by Koopman and von Neumann (KvN) in the Thirties.
In particular we show that the introduction of the KvN Hilbert space 
of complex and square integrable ``wave functions"
requires an enlargement of the set of the observables of ordinary classical mechanics.
The possible role and the meaning of these extra observables
is briefly indicated in this work. We also analyze the similarities and
differences between non selective measurements and two-slit experiments 
in classical and quantum mechanics.
\end{abstract}
\vfil
\end{titlepage}
\newpage

\section{Introduction}

It is well known that in classical statistical mechanics the evolution of the probability
densities in phase space $\rho(q,p,t)$ is given by the Liouville equation
\begin{equation}
\displaystyle i\frac{\partial}{\partial t}\rho(q,p,t)=\hat{\cal H}\rho(q,p,t) \label{Liorho}
\end{equation}
where $\hat{\cal H}$ is the Liouville operator
\begin{equation}
\hat{\cal H}= -i\partial_pH(q,p)\partial_q+i\partial_qH(q,p)\partial_p \label{liouv}
\end{equation}
and $H(q,p)$ is the Hamiltonian in the usual phase space ${\cal M}$. Since the distributions $\rho(q,p)$
are probability densities they must only be {\it integrable} functions: $\rho(q,p)\in L^1$. In 
a previous paper \cite{waves1} we stressed that KvN did not use the space 
of the $\rho$ but introduced instead a Hilbert space made up
of complex {\it square integrable} functions $\psi(q,p)\in L^2$ over phase space \cite{Koopman}\cite{von Neumann}.
These $\psi$ are ``the KvN waves" we indicated in the title. Next they {\it postulated} 
for every $\psi(q,p,t)$ an
equation of evolution which is the Liouville equation itself:
\begin{equation}
\displaystyle 
i\frac{\partial}{\partial t}\psi(q,p,t)=\hat{\cal H}\psi(q,p,t). \label{Liopsi}
\end{equation}
Because the Liouvillian $\hat{\cal H}$ contains only first order derivatives, it is  
easy to prove that the Liouville equation (\ref{Liorho}) for the probability densities
$\rho(q,p,t)$ can be derived from (\ref{Liopsi}) by postulating that $\rho(q,p,t)=|\psi(q,p,t)|^2$.

Every theory formulated via states and operators in a Hilbert space can be formulated 
also via path integrals. It has been proved in Ref. \cite{gozzi} 
that one can describe the evolution (\ref{Liopsi}) by means of the
following kernel of propagation:
\begin{equation}
\displaystyle \psi(\varphi,t)=\int d\varphi\, K(\varphi,t|\varphi_i,t_i)\psi(\varphi_i,t_i) \label{evoker}
\end{equation}
where we have indicated with $\varphi\equiv (q,p)$ all the phase space variables. The kernel
of propagation $K(\varphi,t|\varphi_i,t_i)$ of (\ref{evoker}) can be given the following path integral expression
\cite{waves1}:
\begin{equation}
\displaystyle K(\varphi,t|\varphi_i,t_i)=\int {\cal D}^{\prime\prime}
\varphi{\cal D}\lambda \,\textrm{exp}\biggl[i\int dt\, {\cal L}\biggr] \label{path}
\end{equation}
where ${\cal L}$ is the following Lagrangian: 
${\cal L}=\lambda_a\dot{\varphi}^a-{\cal H}$ with ${\cal H}$ given by 
\begin{equation}
{\cal H}=\lambda_a\omega^{ab}\partial_bH\equiv \lambda_q\partial_pH-\lambda_p\partial_qH \label{liouvilliano}.
\end{equation}
The symbol 
${\cal D}^{\prime\prime}$ in (\ref{path}) indicates 
that the integration is over paths with fixed end points in $\varphi$. 
Having a path integral we can introduce the concept of a commutator as Feynman
did in the quantum case. It goes as follows: given two functions $O_{\scriptscriptstyle 1}(\varphi,\lambda)$ and
$O_{\scriptscriptstyle 2}(\varphi,\lambda)$ the commutator is defined as:
\begin{equation}
\langle[O_{\scriptscriptstyle 1},O_{\scriptscriptstyle 2}]\rangle\equiv\lim_{\epsilon\to 0}\langle
O_{\scriptscriptstyle 1}(t+\epsilon)O_{\scriptscriptstyle 2}(t)-O_{\scriptscriptstyle 2}
(t+\epsilon)O_{\scriptscriptstyle 1}(t)\rangle
\end{equation}
where $\langle\quad\rangle$ means the expectation value under the path integral (\ref{path}).
What we get is \cite{gozzi}:
\begin{equation}
[\varphi^a,\varphi^b]=0,\qquad\quad [\lambda_a,\lambda_b]=0,
\qquad\quad [\varphi^a,\lambda_b]=i\delta_b^a. \label{comm}
\end{equation}
The first commutator of (\ref{comm}) tells us that the positions $q$ commute with the momenta $p$, confirming 
that we are doing classical and not quantum mechanics. 
The last commutator instead tells us the $\lambda_a$ are ``something like"
the momenta conjugate to $\varphi^a$. In order to satisfy (\ref{comm}) we can use the following representation:
\begin{equation}
\displaystyle \hat{\varphi}^a=\varphi^a,\qquad \hat{\lambda}_a=-i\frac{\partial}{\partial \varphi^a}.
\end{equation}
Via the previous operatorial realization, the ${\cal H}$ of (\ref{liouvilliano}) can be turned
into an operator:
\begin{equation}
\label{liouvillianoop}
{\cal H}\,\rightarrow\, \hat{\cal H}=-i\omega^{ab}\partial_bH\partial_a=-i\partial_pH\partial_q+i\partial_qH\partial_p.
\end{equation}
Therefore $\hat{\cal H}$ is just the Liouville operator of Eq. (\ref{liouv}). This confirms that the operatorial
formalism generated by the path integral (\ref{path}) is nothing else than the KvN one. The careful
reader may have noticed that in this paper we have neglected the introduction of those auxiliary Grassmann
variables $c^a, \bar{c}_a$ which made their appearance in \cite{gozzi}.
The reason is that they are not necessary for the issues we are going to discuss in this paper.
Technically we could say that we restrict ourselves to the zero-form number sector 
\cite{gozzi} of the full Hilbert space studied in \cite{2bis}.
 
In Ref. \cite{waves1} we proved that the presence of only first order derivatives in the Liouvillian
has a lot of peculiar consequences for what concerns the role of phases in the KvN theory.
For example, differently from what happens in quantum mechanics (QM), 
the equations of motion for the phase and the modulus of the $\psi(\varphi)$ are completely 
decoupled. Actually it is in the particular {\it representation} in which both the positions and the momenta
are given by multiplicative operators that the phase does not enter the equations of motion of 
the modulus and vice versa. Not only, but in such a representation the phases do not bring 
any information about the mean values of the observables $O(\hat{\varphi})$.
As a consequence one could think that phases are completely useless in CM. 
It is not so; in fact we proved in Ref. \cite{waves1} that, if we want to remain free to consider 
also other representations for $\hat{\varphi}$ and $\hat{\lambda}$, then we have to use 
complex ``wave functions" since their phases are then related to the mean values of the usual
observables $O(\hat{\varphi})$ of CM. We can say that in \cite{waves1}, in order
to understand whether the phases of the KvN ``wave functions" have or not a physical meaning,
we concentrated ourselves on the role played by the different {\it representations} of the Hilbert space of CM.
In this paper instead we will stick ourselves to the representation in which the KvN ``wave functions"
are given by $\psi(\varphi)$ and we will analyze in detail the role played by the different
choice of {\it observables} within the KvN formalism.

In particular a topic which was not analyzed in the previous paper \cite{waves1} on the KvN waves 
is the following: if we consider a Hilbert space of {\it complex} elements $\psi(\varphi)$ 
then we must enlarge the set of observables 
of the theory. In fact if we stick to the accepted wisdom that in classical mechanics (CM) the 
observables are only the functions of $\varphi$ then a {\it superselection} mechanism
sets in and it forces the Hilbert space to be just the set of all Dirac deltas centered 
on single points of phase space. In order to give a physical meaning to $\psi(\varphi)$,
which are linear superpositions with complex coefficients of the Dirac deltas mentioned above,
we have to prevent the superselection mechanism from acting and this can be done by enlarging 
the space of ``observables" to functions not only of $\varphi$ but also of $\partial/\partial\varphi$. 
In this way it becomes meaningful
to build a Hilbert space made up of square integrable functions with phases. 
In fact the phases of the KvN functions $\psi(\varphi)$ could 
in principle be measured
by using ``observables" depending not only on $\hat{\varphi}$ but also on $\partial/\partial\varphi$.
It would be interesting to clarify and understand the physical meaning 
of these ``observables". A typical operator of this kind, which depends on both 
$\varphi$ and $\partial/\partial\varphi$,
is the Liouvillian $\hat{\cal H}$ of (\ref{liouv}). It is well known
\cite{Arnold} that from its spectrum we can extract 
physical information regarding, for example, the {\it ergodicity} of the system and, if we extend 
$\hat{\cal H}$ to the Lie derivative along the Hamiltonian flow ${\cal L}_h$
\cite{gozzi}, we can even extract 
information on concepts like {\it Lyapunov exponents} and {\it entropies} 
\cite{3bis}. More on this issue can be found in {\bf Sec. 2}. 

Another topic which was missing in Ref. \cite{waves1} is the following:
if we want to extract 
information on the properties of the KvN states by means of suitable measurements of observables, 
then we have first to clarify 
which is the effect of a process of {\it measurement} in the KvN formalism. 
In {\bf Sec. 3} we impose, on the operatorial approach to CM,
the same postulates about measurements as those of standard QM. Nevertheless, 
the fact that the KvN commutators among canonical 
variables and their equations of motion are different from those in QM   
create crucial differences between the two theories 
as we will study in detail in the particular case of {\it non selective 
measurements}. In CM a non selective measurement of $\hat{\varphi}$ does not change the probability 
distributions of $\varphi$ either immediately after the measurement or after a long time.
In QM instead a non selective measurement of $\hat{x}$ disturbs immediately the probability
distribution in $p$ and it changes
also the probability distribution in $x$ during the time evolution,
since $x$ and $p$ are coupled in the equations of motion. 

A similar analysis is used in {\bf Sec. 4} to explain what happens in the two-slit experiment:
in CM the presence of the screen with the two slits does not create an interference figure
in the $\varphi$ variables at all. In QM instead the two slits create immediately an 
interference figure in the distribution of the momenta $p$. Since $x$ and $p$ are coupled in the
equations of motion, this interference figure propagates itself also to the coordinates $x$
during the time evolution and this produces the well known quantum interference effects. 
In {\bf Sec. 5} we give some brief conclusions.

\section{Superselection Rules and Observables}

As we said in the Introduction, in the KvN formulation of CM the Hilbert space
is made up of {\it complex} ``wave functions" over the phase space variables. 
In this section we want first to analyze in detail which assumptions have to be done on the
observables of the theory in order to justify the choice 
of complex ``wave functions" mentioned above. Second we want to give the abstract
theoretical reasons why the phases of the classical KvN waves cannot be detected by those operators 
which depend only on $\hat{\varphi}=(\hat{q},\hat{p})$. 
In performing this analysis
we shall use the notion of superselection rules. For a review about this subject we refer the reader to
\cite{mex}\cite{gal}.

In the standard formulation of CM via phase space and Poisson brackets, 
the observables are usually identified with the real functions 
of the phase space variables $\varphi$. In the KvN formulation of CM this is equivalent 
to postulate that the observables are {\it all} and {\it only} the Hermitian functions of the operators
$\hat{\varphi}=(\hat{q},\hat{p})$. Let us accept for a while this postulate and see which 
are its consequences. 
First of all the algebra of the classical observables turns out to be Abelian and the operators
$\hat{\varphi}=(\hat{q},\hat{p})$ commute with {\it all} the observables 
of the theory. Therefore the operators $\hat{\varphi}=(\hat{q},\hat{p})$
are {\it superselection operators} and the associated superselection rules have to be taken into account. According to
these rules if
we consider two states corresponding to different eigenvalues of the superselection operators, i.e. 
$|\varphi_{\scriptscriptstyle 1}\rangle$ and $|\varphi_{\scriptscriptstyle 2}\rangle$ satisfying:
\begin{equation}
\left\{
\begin{array}{l}
\hat{\varphi}|\varphi_{\scriptscriptstyle 1}\rangle=\varphi_{\scriptscriptstyle 1}|\varphi_{\scriptscriptstyle 1}
\rangle \smallskip \\
\hat{\varphi}|\varphi_{\scriptscriptstyle 2}\rangle=\varphi_{\scriptscriptstyle 2}|\varphi_{\scriptscriptstyle 2}
\rangle,
\end{array}
\right.
\end{equation} 
we have that there is no
observable connecting them since:
\begin{equation}
\langle\varphi_{\scriptscriptstyle 1}|O(\hat{\varphi})|\varphi_{\scriptscriptstyle 2}\rangle=0.
\end{equation}
Furthermore if we consider a linear superposition of eigenstates corresponding to different eigenvalues:
\begin{equation}
|\psi\rangle=\alpha_{\scriptscriptstyle 1}|\varphi_{\scriptscriptstyle 1}\rangle
+\alpha_{\scriptscriptstyle 2}|\varphi_{\scriptscriptstyle 2}\rangle \label{alpha}
\end{equation}
with $\alpha_{\scriptscriptstyle 1}\alpha_{\scriptscriptstyle 2}\neq 0$ we have that the vector 
$|\psi\rangle$ cannot be considered a {\it pure} state. 
In fact a superposition like the $|\psi\rangle$ of (\ref{alpha})
cannot be the eigenstate of any observable of the form $O(\hat{\varphi})$. So, if we limit the observables to the 
Hermitian functions of $\hat{\varphi}$, then the state $|\psi\rangle$ cannot be prepared 
by diagonalizing a complete set of observables, like one does 
for all the pure states of QM.
Besides this when we compute the expectation values of the observables $O(\hat{\varphi})$ 
on $|\psi\rangle$ we have that:
\begin{equation}
\displaystyle \overline{O}=
\frac{\langle\psi|O(\hat{\varphi})|\psi\rangle}
{\langle \psi|\psi\rangle}=
\frac{|\alpha_{\scriptscriptstyle 1}|^2\langle\varphi_{\scriptscriptstyle 1}|
O(\hat{\varphi})|\varphi_{\scriptscriptstyle 1}\rangle+|\alpha_{\scriptscriptstyle 2}|^2
\langle\varphi_{\scriptscriptstyle 2}|O(\hat{\varphi})|\varphi_{\scriptscriptstyle 2}\rangle}
{|\alpha_{\scriptscriptstyle 1}|^2\langle\varphi_{\scriptscriptstyle 1}|\varphi_{\scriptscriptstyle 1}\rangle
+|\alpha_{\scriptscriptstyle 2}|^2\langle\varphi_{\scriptscriptstyle 2}|\varphi_{\scriptscriptstyle 2}\rangle}. 
\label{gamma}
\end{equation}
Therefore the expectation values that can be calculated using the vector $|\psi\rangle$
are just the same as those calculated starting from the {\it mixed} density matrix
\begin{equation}
\hat{\rho}=|\alpha_{\scriptscriptstyle 1}|^2|\varphi_{\scriptscriptstyle 1}\rangle\langle
\varphi_{\scriptscriptstyle 1}|
+|\alpha_{\scriptscriptstyle 2}|^2|\varphi_{\scriptscriptstyle 2}\rangle\langle\varphi_{\scriptscriptstyle 2}|
\label{beta}
\end{equation}
via the rule $\displaystyle
\overline{O}=\textrm{Tr}[\hat{\rho}\;O(\hat{\varphi})]/\textrm{Tr}[\hat{\rho}]$.
So from a physical point of view the state (\ref{alpha}) cannot be distinguished from 
the mixed density matrix (\ref{beta}). This means that ``{\it coherent superpositions of pure states are impossible,
one automatically gets mixed states when attempting to form them}" \cite{lan}. 
All this can be
rephrased in the following way: the relative phase between $|\varphi_{\scriptscriptstyle 1}\rangle$
and $|\varphi_{\scriptscriptstyle 2}\rangle$ cannot be measured using only observables like $O(\hat{\varphi})$.
In fact for the mean values of these observables the only crucial element  
is the modulus square of $\alpha_{\scriptscriptstyle
1}$ and $\alpha_{\scriptscriptstyle 2}$ as it is clear from (\ref{gamma}).

These considerations can be extended very easily to the case of a continuous superposition of states 
$|\varphi\rangle$. In fact the two vectors 
\begin{equation}
\displaystyle |\psi\rangle=\int d\varphi\,\psi(\varphi)|\varphi\rangle,\qquad\qquad
|\widetilde{\psi}\rangle=\int d\varphi\,\psi(\varphi)e^{iA(\varphi)}|\varphi\rangle \label{lialia}
\end{equation}
are physically indistinguishable because they give the same expectation values for all the observables 
of the form $O(\hat{\varphi})$. 
``{\it But they are also completely different vectors in Hilbert space!$\,$}" \cite{mex}. 
If we want to avoid 
the redundancy represented by the states $|\psi\rangle$ and $|\widetilde{\psi}\rangle$
of (\ref{lialia}), the only way is to forbid the superposition of eigenstates of the superselection operators
and consider only the statistical mixtures (\ref{beta}) which, in the continuous case, become:
\begin{equation}
\displaystyle \hat{\rho}=\int d\varphi\,\rho(\varphi)|\varphi\rangle\langle \varphi|. \label{delta}
\end{equation}
Therefore the Hilbert space must be considered as a direct sum (or, better to say, a direct integral)
\begin{equation}
\displaystyle {\mathbf{H}}=\oplus_{\scriptscriptstyle
\{\varphi_i\}}{\mathbf{H}}_{\{\varphi_i\}}
\end{equation}
of the different eigenspaces ${\mathbf{H}}_{\{\varphi_i\}}$ corresponding to the different 
eigenvalues $\varphi_i$ 
for the superselection observables
$\hat{\varphi}$. These eigenspaces are incoherent, i.e. the relative phases between vectors
belonging to different eigenspaces cannot be measured at all; not only, but it is impossible to move from one
eigenspace to the other by means of an observable. 

In QM the superselection observables, like the parity or the charge operator, commute with
all the observables of the system and in particular with the Hamiltonian $\hat{H}$, which is the generator 
of the time evolution. This implies that the eigenvalues of the superselection operators
are constants of the motion. So when we prepare the system in one particular eigenspace the time evolution
cannot bring the system outside of it. This is fine if the superposition operator is the parity or the
charge because it only implies that all the states reached by the time evolution have the same parity
and the same charge. But this is catastrophic in the KvN formulation of CM. In fact there the superselection
operators are the $\hat{\varphi}$ and so the eigenspaces of the superselection operators are in 1-1
correspondence with the points of the phase space ${\cal M}$; when we consider the time evolution of the system 
we move from one point of the phase space to the other and, therefore, from one eigenspace of the superselection 
operators to the other. Therefore if the observables of CM are only the functions of 
$\hat{\varphi}$ and as a consequence the superselection mechanism is automatically triggered, then we have to admit 
that the time evolution of the system cannot be performed by an operator belonging to the observables of
the system.
We note that this is perfectly consistent with the fact that the generator of the evolution is
the Liouvillian which depends also on $\hat{\lambda}$ and not only on $\hat{\varphi}$,
see (\ref{liouvilliano}) and (\ref{liouvillianoop}).

There are two basic scenarios that we can envision. In the first one 
we can insist in having as observables only the Hermitian functions
$O(\hat{\varphi})$ and as a consequence in considering as physically significant only the statistical mixtures
(\ref{delta}). From the probability density $\rho(\varphi)$ we can then construct its {\it real} square root
$\psi(\varphi)\equiv |\sqrt{\rho(\varphi)}|$ and use it to build the following 
linear superposition of the eigenstates $|\varphi\rangle$:
\begin{equation}
\displaystyle |\psi\rangle=\int d\varphi\, \psi(\varphi)|\varphi\rangle. \label{epsilon}
\end{equation}
Now the coefficients $\psi(\varphi)$ in the superposition (\ref{epsilon}) 
are just {\it real} functions of $\varphi$.
So there is a 1-1 correspondence between the statistical mixtures (\ref{delta}) and the 
vectors $|\psi\rangle$ of (\ref{epsilon}). Not only, let us construct from the $|\psi\rangle$
of (\ref{epsilon})
the following pure density matrix 
\begin{equation}
\hat{\rho}^{\,\prime}=|\psi\rangle\langle \psi|=\int d\varphi d\varphi^{\prime}\,\psi(\varphi)\psi(\varphi^{\prime})
|\varphi\rangle\langle \varphi^{\prime}|. \label{delta2}
\end{equation}
Since the algebra of the observables $O(\hat{\varphi})$ is Abelian we have that only the diagonal terms 
of $\hat{\rho}^{\,\prime}$ contribute to the mean values of $O$. As a consequence it is easy to realize 
that the mean values of the observables calculated from the $\hat{\rho}$ 
of (\ref{delta}) and the $\hat{\rho}^{\,\prime}$ of (\ref{delta2}) are just the same:
\begin{equation}
\displaystyle \frac{\textrm{Tr}[\hat{\rho}\,O(\hat{\varphi})]}{\textrm{Tr}[\hat{\rho}]}=
\frac{\textrm{Tr}[\hat{\rho}^{\,\prime}
O(\hat{\varphi})]}{\textrm{Tr}[\hat{\rho}^{\,\prime}]}.
\end{equation}
So we can say that with (\ref{epsilon}), i.e. a linear superposition of $|\varphi\rangle$ with {\it real}
coefficients, we can perform the same physics as with the statistical mixture $\hat{\rho}$ of
(\ref{delta}). Furthermore there is a 1-1 correspondence between the {\it real} vectors $|\psi\rangle$
and the statistical mixtures $\hat{\rho}$. So there is no redundancy in the description and we can 
look at the states $|\psi\rangle$ just as useful mathematical tools to perform all the calculations
and make all the predictions of classical statistical mechanics. Since the coefficients of the superposition
(\ref{epsilon}) are real we have that the operators depending on $\hat{\varphi}$ cannot feel the phases  
just because there is no phase at all.

The second scenario is to take as observables all the Hermitian operators of the KvN theory.
With this assumption also
the operators $\hat{\lambda}$ and their Hermitian functions become observables.
Now we can no longer say that $\hat{\varphi}$ commute with all the observables
since $[\hat{\varphi}^a,\hat{\lambda}_b]=i\delta_b^a$. As a consequence the superselection mechanism is not
triggered
and a coherent superposition, even with complex coefficients, 
of the eigenstates $|\varphi\rangle$ can be prepared by diagonalizing 
a complete set of commuting operators. This is the solution that also KvN must have had in mind\footnote{We note that the
first paper on superselection rules made its appearance
only in 1952 \cite{wick}, more or less twenty years after KvN's papers \cite{Koopman}\cite{von Neumann}.}
in their original papers when they
{\it postulated} that the elements of their Hilbert space were square integrable and {\it complex} functions.
In fact, from what we said in the first scenario, it would not be necessary to have {\it complex}
wave functions if we restricted the set of observables to be only the functions $O(\hat{\varphi})$.
Insisting in having complex wave functions and no degeneracy like the one given by the 
$|\psi\rangle$ and $|\widetilde{\psi}\rangle$ of (\ref{lialia}), is equivalent to saying that the 
set of observables is not just the $O(\hat{\varphi})$ but the whole set of Hermitian operators 
$O(\hat{\varphi},\hat{\lambda})$. 
This is the solution that we also have implicitly adopted in Ref. \cite{waves1}. 
Such a theory is a {\it generalization}
of standard CM since it contains a much larger set of observables. 

It would be interesting
to understand the physical meaning of these extra observables depending on $\hat{\lambda}$. 
One of these ``observables" is the Liouvillian $\hat{\cal H}=\hat{\lambda}_a\omega^{ab}\partial_bH$
and we know that its spectrum gives us information
on such properties as the {\it ergodicity} of the system. It is in fact well known \cite{Arnold}
that if the zero eigenvalue of $\hat{\cal H}$ is non degenerate then the system is ergodic. 
From the generalization of $\hat{\cal H}$ to the space of forms \cite{gozzi} we get 
what is known as the Lie derivative of the Hamiltonian flow ${\cal L}_h$. It was proved in Ref. \cite{3bis}
that from the spectrum of this operator one could derive information 
on the Lyapunov exponents and the related entropies. Anyhow all these information 
(ergodicity, Lyapunov exponents,\ldots) obtained from the spectra of the Liouvillian
$\widetilde{\cal H}$ or the Lie derivative ${\cal L}_h$
could also be derived using only the Hamiltonian 
$H(\varphi)$ and the correlations of $\hat{\varphi}$ at different times. 
We should note anyhow that in the KvN formalism there are plenty of other Hermitian operators,
besides the Liouvillian $\hat{\cal H}$, which contain the variables $\hat{\lambda}$.
Actually it has been proved recently by one of us \cite{quantum} that among these generalized
``observables" of the KvN theory there is a set which is exactly isomorphic to the standard 
observables of QM. This gives us hope that the KvN formalism may be the 
right tool to use in order to explore that tricky region which is at the border 
between CM and QM \cite{Leggett}\cite{grw}. In that region there may be phenomena 
which need coherent superpositions of states like the one given in (\ref{alpha}). 

To conclude this section 
we can say that the KvN theory is a generalization of CM once we admit
among the observables also the Hermitian operators containing $\hat{\lambda}$.
It reduces to CM if we restrict the observables to be only the Hermitian operators made out of $\hat{\varphi}$.
Using the arguments explained above via the superselection principle it is possible
to prove that these last observables could never 
detect the relative phases contained in the coherent superposition (\ref{alpha}). 

\section{Measurements in QM and CM} 

Another aspect which is worth investigating in order to get a better
understanding of the differences and the similarities between CM and QM is 
related to the measurement issue. We know in fact that in QM a measurement
disturbs the system and modifies the probability of the 
outcomes of subsequent measurements. This disturbance is present 
even if the measurement is a {\it non selective} one, i.e. 
a measurement
which is explicitly performed but whose results are not read out.
We will see later on that in this case the system does not ``collapse" on a particular eigenstate 
but on a incoherent superposition of all its possible eigenstates. 
We want to begin this analysis with a very simple but pedagogical exercise \cite{Ghirardi}:

\subsection{Effect of non Selective Measurements in QM}

Consider a {\it quantum} mechanical system characterized by a Hermitian Hamiltonian
with eigenvalues $E_{\scriptscriptstyle 1}=\hbar\omega$; $E_{\scriptscriptstyle 2}=-\hbar\omega$
and eigenfunctions $|+\rangle$ and $|-\rangle$ respectively. Let us take also an observable 
$\hat{\Omega}$ with eigenvectors:
\begin{equation}
\displaystyle
|a\rangle=\frac{1}{\sqrt{2}}\Bigl[|+\rangle+|-\rangle\Bigr],\qquad\quad
|b\rangle=\frac{1}{\sqrt{2}}\Bigl[|+\rangle-|-\rangle\Bigr].
\end{equation}
Suppose we prepare the system in the following initial state:
\begin{equation}
\displaystyle |\psi,0\rangle=\frac{1}{2}|+\rangle+\sqrt{\frac{3}{4}}|-\rangle.
\end{equation}

\begin{itemize}
\item[{\bf 1)}] Let us evolve the system up to time $t=2\tau$ and calculate the probability
of obtaining $a$ as result of a measurement of $\hat{\Omega}$. The wave function at time $t=2\tau$
is given by:
\begin{equation}
\displaystyle |\psi,2\tau\rangle=e^{-\frac{i}{\hbar}\hat{H}2\tau}\Biggl[\frac{1}{2}|+\rangle
+\sqrt{\frac{3}{4}}|-\rangle\Biggr]=\frac{1}{2}|+\rangle e^{-2i\omega\tau}+\sqrt{\frac{3}{4}}|-\rangle
e^{2i\omega\tau}. \label{purestate}
\end{equation}
So, when the system is described by the {\it Pure state} (\ref{purestate}), 
the probability of finding $a$ as result of a measurement of $\hat{\Omega}$ at time $t=2\tau$ is given by
\begin{equation}
P_{\scriptscriptstyle P}(a|2\tau)=\Bigl| \langle \psi,2\tau|a\rangle\Bigr|^2=
\frac{1}{2}\Bigl(1+\sqrt{\frac{3}{4}}\textrm{cos}\,4\omega\tau
\Bigr). \label{vercinge}
\end{equation}
\item[{\bf 2)}] Let us now perform at time $t=\tau$ a {\it non selective} measurement (NSM)
of the observable $\hat{\Omega}$.
Are the probabilities at time $t=2\tau$ influenced by the fact that we have measured $\hat{\Omega}$
at time $t=\tau$ but not read the result? To answer this question let us consider what happens just after the measurement. Because 
of the postulate of the collapse of the wave function and because of the fact that in a NSM 
we do not read out the result of the measurement, 
the system will be described by a statistical mixture of the two
eigenstates $|a\rangle$ and $|b\rangle$ of $\hat{\Omega}$ where the weights are just the probabilities 
$P(a|\tau)$ and $P(b|\tau)$ that the results $a$ and $b$ are obtained at time $t=\tau$:
\begin{equation}
\hat{\rho}_{\scriptscriptstyle M}(\tau)=P(a|\tau)|a\rangle
\langle a|+P(b|\tau)|b\rangle\langle b|.
\end{equation}
The index $M$ has been put on $\hat{\rho}$ to indicate that $\hat{\rho}_{\scriptscriptstyle M}$ 
is a {\it Mixed} density matrix.
Now let the system evolve up to time 
$t=2\tau$. What we obtain is:
\begin{equation}
\hat{\rho}_{\scriptscriptstyle M}(2\tau)=P(a|\tau)|\psi_a,2\tau\rangle
\langle \psi_a,2\tau|+P(b|\tau)|\psi_b,2\tau\rangle\langle \psi_b,2\tau|
\end{equation}
where $|\psi_a,2\tau\rangle$ and $|\psi_b,2\tau\rangle$ are given by the evolution of
the eigenstates $|a\rangle$ and $|b\rangle$ from $t=\tau$ to
$t=2\tau$:
\begin{equation}
\left\{
\begin{array}{l}
\displaystyle |\psi_a,2\tau\rangle=e^{-\frac{i}{\hbar}\hat{H}\tau}|a\rangle=
\frac{1}{\sqrt{2}}\Bigl(e^{-i\omega\tau}|+\rangle+e^{i\omega\tau}|-\rangle\Bigr),\medskip\\
\displaystyle |\psi_b,2\tau\rangle=e^{-\frac{i}{\hbar}\hat{H}\tau}|b\rangle=
\frac{1}{\sqrt{2}}\Bigl(e^{-i\omega\tau}|+\rangle-e^{i\omega\tau}|-\rangle\Bigr).
\end{array}
\right.
\end{equation}
Let us now calculate the probability of obtaining $a$ as result of a further measurement 
of $\hat{\Omega}$ at time $t=2\tau$. We have to calculate:
\begin{equation}
P_{\scriptscriptstyle M}(a|2\tau)=\textrm{Tr}\Bigl[\hat{\rho}_{\scriptscriptstyle M}(2\tau)|a\rangle
\langle a|\Bigr]
\end{equation}
whose result is given by 
\begin{equation}
\displaystyle P_{\scriptscriptstyle M}(a|2\tau)=\frac{1}{2}\Bigl(1+\sqrt{\frac{3}{4}}\textrm{cos}^2
2\omega\tau\Bigr).
\end{equation}
Note that this result is different than the one obtained in (\ref{vercinge}). 
\end{itemize}

So we can conclude this 
exercise by saying that, if a NSM 
of $\hat{\Omega}$ is performed at time $t=\tau$ like in the case {\bf 2)},
then the probabilities of the outcomes 
of $\hat{\Omega}$ itself at time $t=2\tau$ are {\it modified} with respect to the case {\bf 1)}
in which such a measurement is not performed. 
This can be also summarized in the following scheme:

\bigskip
\bigskip

\begin{center}
{\Large{$\boxed{\;\textrm{Initial Wave function} \; |\psi,0\rangle}\quad $}} 

{\large{
$\begin{array}{cc}
\quad \Big|  \qquad  & \qquad \qquad  \Big| \\
\quad \Big|  \qquad  & \qquad \qquad  \Big| \\
\quad \textrm{case} \;{\bf 1)} \qquad  & \qquad \qquad {\bf 2)} \; \textrm{NSM \;of\;} \hat{\Omega} \\
\quad |\psi,\tau\rangle       \qquad  &  \qquad \qquad  
|\psi,\tau\rangle\rightarrow \hat{\rho}_{\scriptscriptstyle M}(\tau)\\
\quad \Big|  \qquad  &  \qquad \qquad  \Big|\\
\quad \Big\downarrow \qquad & \qquad \qquad \Big\downarrow \medskip\\
\quad |\psi,2\tau\rangle \qquad & \qquad \qquad \hat{\rho}_{\scriptscriptstyle M}(2\tau)
\end{array}$}}

\hspace{-0.5cm} {\large Outcome 1 for $\hat{\Omega}$ \qquad $\neq$ \qquad Outcome 2 for $\hat{\Omega}$}
\end{center}

\bigskip

\subsection{Non Selective Measurements of $\hat{x}$ in QM}

In order to make a more direct comparison between classical and quantum mechanics in this subsection
we want to analyze what happens in QM when we perform at a
certain time a NSM of an operator with a continuous spectrum like $\hat{x}$. To be as clear as possible
let us distinguish the following cases:

\smallskip

\begin{itemize}
\item[{\bf a)}] the case in which we do not perform any measurement of $\hat{x}$ at time $t=0$, we let the system
evolve and we calculate the probabilities of the outcomes of a measurement of $\hat{x}$ at time $t=\tau$.

\smallskip

\item[{\bf b)}] the case in which we perform a NSM of $\hat{x}$ at $t=0$ and then

\smallskip

\begin{itemize}
\item[{\bf b1)}] either we perform another measurement (of $\hat{x}$ or $\hat{p}$) 
just after the first one at 
$t=0_+$ (we will use the notation $0_+$ in order to indicate that the second measurement is performed 
again at $t=0$ but {\it after} the first one) 

\smallskip

\item[{\bf b2)}] or we let the system evolve from $t=0$ to a finite $t=\tau$ and
we calculate at that time the probabilities of the outcomes of a measurement of $\hat{x}$.
\end{itemize}
\end{itemize}

\smallskip

\begin{itemize}
\item[{\bf a)}] Let us start with a quantum wave function $|\psi,0\rangle$; performing its evolution
up to time $t=\tau$, we obtain another state $|\psi,\tau\rangle$. Let us then ask ourselves: which is
the probability density of finding a particle between $x$ and $x+dx$ 
as result of a measurement of $\hat{x}$
at time $t=\tau$? The answer is:
\begin{equation}
\rho_{\scriptscriptstyle P}(x|\tau)=\textrm{Tr}\Bigl[|x\rangle\langle x
|\hat{\rho}_{\scriptscriptstyle P}(\tau)\Bigr]
=|\psi(x,\tau)|^2.
\end{equation}
The index $P$ indicates that the matrix $\hat{\rho}_{\scriptscriptstyle P}(\tau)=
|\psi,\tau\rangle\langle \psi,\tau|$ is a {\it Pure} one.
For example if we consider a {\it free} particle described by the following Gaussian wave function:
\begin{equation}
\displaystyle
\psi(x,0)=\frac{1}{\sqrt{\sqrt{\pi}a}}\textrm{exp}\biggl(-\frac{x^2}{2a^2}+\frac{i}{\hbar}p_ix\biggr) \label{init1}
\end{equation}
at time $t=\tau$ we obtain that 
the probability density $\rho_{\scriptscriptstyle P}(x|\tau)$ is given by the modulus square of 
$\psi(x,\tau)$:
\begin{equation}
\displaystyle
\rho_{\scriptscriptstyle P}(x|\tau)= N\cdot
\textrm{exp}\biggl[-\frac{m^2a^2}{m^2a^4+\hbar^2
\tau^2}\biggl(x-\frac{p_i\tau}{m}\biggr)^2\biggr] \label{bip}
\end{equation}
i.e. another Gaussian with a well defined mean value and a finite standard deviation.

\smallskip

\item[{\bf b)}] Let us now perform a NSM
of the quantum position $\hat{x}$ at time $t=0$.
With the initial wave function $|\psi,0\rangle$ the probability density of finding the system 
between $x_0$ and $x_0+dx_0$ as result of a measurement is given by 
\begin{equation}
\rho(x_0|0)=|\psi(x_0,0)|^2.
\end{equation}
Just after the measurement, if the result is not read out, 
the system is described by the statistical mixture 
\begin{equation}
\hat{\rho}_{\scriptscriptstyle M}(0_+)=\int dx_0 |\psi(x_0,0)|^2|x_0\rangle\langle x_0| \label{sm}
\end{equation}
instead of the pure one:
\begin{equation}
\hat{\rho}_{\scriptscriptstyle P}(0_-)=|\psi,0\rangle\langle\psi,0|
\end{equation}
which described the system just before the measurement. From now on we will use the notation
$0_-$ to indicate that we are at $t=0$ but {\it before} the measurement.
The question we want to answer is: has the NSM 
of $\hat{x}$ changed the probability distributions
of the observables? 

\smallskip

\item[{\bf b1)}] First of all, without taking into account any time evolution, let us perform 
a further measurement at $t=0_+$. 
If the observable we measure is again the position operator $\hat{x}$, than  
the probability distribution of the outcomes is left unchanged.
In fact before the measurement done at $t=0$ we had the following 
distribution in $x$:
\begin{equation}
\rho_{\scriptscriptstyle P}(x|0_-)=
\textrm{Tr}\Bigl[|x\rangle\langle x|\hat{\rho}_{\scriptscriptstyle P}(0_-)\Bigr]
=\psi^*(x,0)\psi(x,0). \label{measbef}
\end{equation}
Note that it remains the same also after the measurement at $t=0$:
\begin{equation}
\displaystyle \rho_{\scriptscriptstyle M}(x|0_+)=\frac{\textrm{Tr}\Bigl[|x\rangle\langle x|
\hat{\rho}_{\scriptscriptstyle M}(0_+)\Bigr]}{\textrm{Tr}\Bigl[\hat{\rho}_{\scriptscriptstyle M}
(0_+)\Bigr]}=\psi^*(x,0)\psi(x,0). \label{measaft}
\end{equation}
Let us now look at observables different from $\hat{x}$, like for example 
the momentum $\hat{p}$. 
The NSM of $\hat{x}$ at $t=0$ changes immediately the probability density
of finding a particular outcome in a measurement of $\hat{p}$.
In fact before the NSM of $\hat{x}$ at time $t=0$ the distribution in the momentum 
space was given by:
\begin{equation}
\rho_{\scriptscriptstyle P}(p|0_-)=
\textrm{Tr}\Bigl[|p\rangle\langle p|\hat{\rho}_{\scriptscriptstyle P}(0_-)\Bigr]=
\overline{\psi}^*(p,0)\overline{\psi}(p,0). \label{before}
\end{equation}
So, for example, if we consider as initial wave function the Gaussian $\psi(x,0)$ of (\ref{init1}),
its Fourier transform $\overline{\psi}(p,0)$ will be another Gaussian with a well defined mean value $p=p_i$ 
and a finite standard deviation. After the non selective measurement of $\hat{x}$ at $t=0$ the probability
distribution in $p$ will be given by:
\begin{equation}
\displaystyle \rho_{\scriptscriptstyle M}(p|0_+)=
\frac{\textrm{Tr}\Bigl[|p\rangle\langle p|\hat{\rho}_{\scriptscriptstyle M}(0_+)\Bigr]}
{\textrm{Tr} \Bigl[\hat{\rho}_{\scriptscriptstyle M}(0_+)\Bigr]}\textrm{\;\;independent \,of \;}p \label{166}
\end{equation}
and, if we perform explicitly the calculation, it is easy to realize that this probability density is completely 
independent of the momentum $p$. This can be easily understood also without any calculation:
in fact after the measurement at $t=0$ the system is described by a superposition of the eigenstates $|x_0\rangle$ 
like in (\ref{sm}) but, according 
to Heisenberg's uncertainty principle, every eigenstate of the position $|x_0\rangle$ corresponds to a 
completely uniform probability distribution in the momenta. So after the measurement at $t=0$ the overall 
distribution $\rho_{\scriptscriptstyle M}(p|0_+)$ of (\ref{166}) will be different than the one in
(\ref{before}): it will be a uniform distribution, without any mean value and with an infinite 
standard deviation. Therefore we can say that, without taking into account any time evolution, 
the NSM of $\hat{x}$
at the initial time has instantaneously modified the probability distributions of the 
outcomes of the momenta $p$.
The probability of the outcomes of the positions, instead, are left completely unchanged by the NSM
of $\hat{x}$. We can now ask ourselves whether this property is maintained or not during the evolution.

\smallskip

\item[{\bf b2)}] In order to do this let us perform, after the NSM of $\hat{x}$ at time $t=0$,
the evolution of (\ref{sm}). At time $t=\tau$ the system will be described by
the following statistical mixture:
\begin{equation}
\displaystyle \hat{\rho}_{\scriptscriptstyle M}(\tau)=\int dx_0 |\psi(x_0, 0)|^2
|x_0,\tau\rangle\langle x_0,\tau|
\end{equation}
where $|x_0,\tau\rangle$ is given by the evolution of the eigenstate $|x_0\rangle$ up to time $t=\tau$.
If we use the Schr\"odinger representation and the kernel of propagation of a free particle we have that
\begin{equation}
\displaystyle \langle x|x_0,\tau\rangle=\int dx_i\,
\textrm{exp}\biggl[\frac{im}{\hbar\tau}(x-x_i)^2\biggr]\delta(x_i-x_0)=
\textrm{exp}\biggl[\frac{im}{\hbar\tau}(x-x_0)^2\biggr], \label{purephase}
\end{equation}
i.e. a pure phase factor.
If we calculate the probability density of obtaining $x$ as result of a measurement 
of the position $\hat{x}$ at time $\tau$ we obtain:
\begin{equation}
\displaystyle \rho_{\scriptscriptstyle M}(x|\tau)=\frac{\textrm{Tr}\Bigl[|x
\rangle\langle x|\hat{\rho}_{\scriptscriptstyle M}(\tau)\Bigr]}{\textrm{Tr}\Bigl[
\hat{\rho}_{\scriptscriptstyle M}(\tau)\Bigr]}=
\frac{1}{\int_{-\infty}^{\infty}dx}.
\end{equation}
Even from the not so well defined expression above it is easy to understand that there is an equal probability
of finding the particle in any point of the configuration space. Therefore, differently from 
the $\rho_{\scriptscriptstyle P}(x|\tau)$
obtained in the case {\bf a)} in (\ref{bip}),
the probability distribution $\rho_{\scriptscriptstyle M}(x|\tau)$ is uniform.
So the NSM of the position $\hat{x}$ at time $t=0$ 
not only has changed immediately the probability distribution of the momenta $p$, 
as we have analyzed in {\bf b1)}, but it has changed also 
the probability distribution of $x$ at any time $t>0$. 
Now the reader may wonder why the probability distributions
$\rho_{\scriptscriptstyle P}(x|t)$ 
and $\rho_{\scriptscriptstyle M}(x|t)$  
were the same at $t=0$, see (\ref{measbef})-(\ref{measaft}),
but they are different at any time $t>0$. The explanation
is that during the evolution, which is given by $\dot{x}=p$ and couples $x$ with $p$,
the distributions in $x$ are influenced by the initial distributions in $p$ which, as shown in (\ref{before})
and (\ref{166}), are different in the two cases in which we perform, case {\bf b)}, or not perform, case {\bf a)},
the NSM of $\hat{x}$ at $t=0$. So
the NSM of $\hat{x}$ influences
immediately the distribution of probability of the conjugate variable $p$. Next, since
the momenta $p$ are coupled to $x$ via their equations of motion, the changes in the distribution of $p$ 
are inherited by the distribution of the positions at any instant of time $t>0$. 
\end{itemize}

\smallskip

\subsection{Non Selective Measurements of $\hat{\varphi}$ in CM}

Let us now analyze the case of a NSM of $\hat{\varphi}$ in the KvN approach. 
From a formal point of view the situation is the same as in QM with the
following substitutions: $x\longrightarrow \varphi, \;\; p\longrightarrow \lambda$. 
Therefore we can consider the same three cases analyzed before:

\smallskip

\begin{itemize}
\item[{\bf a)}] let us consider an initial pure state 
\begin{equation}
\hat{\rho}_{\scriptscriptstyle P}(0_-)=|\psi,0\rangle\langle\psi,0|.
\end{equation}
Then we let the system evolve up to time $t=\tau$ and at that time we perform a measurement of $\hat{\varphi}$.
The probability density of the possible outcomes $\varphi$ is given by:
\begin{equation}
\rho_{\scriptscriptstyle P}(\varphi|\tau)=\textrm{Tr}\Bigl[|\varphi\rangle\langle
\varphi|\hat{\rho}_{\scriptscriptstyle P}(\tau)\Bigr]=|\psi(\varphi,\tau)|^2.
\end{equation}
For example if we consider a {\it free} particle described by the following Gaussian wave function: 
\begin{equation}
\psi(\varphi,t=0)=\frac{1}{\sqrt{\pi a b}}\textrm{exp}\biggl(-\frac{q^2}{2a^2}-\frac{(p-p_i)^2}{2b^2}\biggl)
\label{ijmpa.double}
\end{equation}
we obtain that at time $t=\tau$
also $\rho_{\scriptscriptstyle P}(\varphi|\tau)$ is a Gaussian with the following 
mean values and standard deviations:
\begin{equation}
\displaystyle
\bar{q}=\frac{p_i\tau}{m},\;\;\;\;\;\;\;\bar{p}=p_i,\;\;\;\;\;\;\;\overline{(\Delta
q(\tau))^2}=\frac{a^2}{2}+\frac{b^2}{2}\frac{\tau^2}{m^2},\;\;\;\;\;\;\; 
\overline{(\Delta p(\tau))^2}=\frac{b^2}{2}.
\end{equation}

\smallskip

\item[{\bf b)}] Let us instead perform at $t=0$ a simultaneous measurement
of the positions and the
momenta, i.e. a measurement of $\hat{\varphi}$, without reading the result. Like in QM also in the KvN 
theory we have the natural ``postulate" that after a measurement the system collapses in the eigenstate
associated with the eigenvalue that we get. In fact this is a typical feature not only
of QM but of every operator formulation of a theory because of its probabilistic interpretation.
In particular since the measurement of $\hat{\varphi}$ that we performed at $t=0$
is non selective, just after the measurement
the system will be described by an incoherent superposition of the eigenstates $|\varphi_0\rangle$
where the weights are given by the associated probability densities
$|\psi(\varphi_0,0)|^2$:
\begin{equation}
\displaystyle \hat{\rho}_{\scriptscriptstyle M}(0_+)=\int d\varphi_0 |\psi(\varphi_0,0)|^2
|\varphi_0\rangle\langle\varphi_0|. \label{mixin}
\end{equation}

\smallskip

\item[{\bf b1)}] As in QM, it is easy to prove that if, without taking into account any time evolution,
at time $t=0_+$ we perform another measurement of $\hat{\varphi}$ 
the probability distributions of the outcomes
are left unchanged by the initial NSM since:
\begin{equation}
\frac{\textrm{Tr}\Bigl[|\varphi\rangle\langle\varphi|\hat{\rho}_{\scriptscriptstyle M}(0_+)\Bigr]}
{\textrm{Tr}\Bigl[\hat{\rho}_{\scriptscriptstyle M}(0_+)\Bigr]}
=\frac{\textrm{Tr}\Bigl[|\varphi\rangle\langle\varphi|\hat{\rho}_{\scriptscriptstyle P}(0_-)\Bigr]}
{\textrm{Tr}\Bigl[\hat{\rho}_{\scriptscriptstyle P}(0_-)\Bigr]}.
\end{equation}
The situation would change if, instead, we were able to perform a measurement of the conjugate operator $\hat{\lambda}$.
In this case, before the NSM of $\hat{\varphi}$ at time $t=0$,
the probability distribution would be:
\begin{equation}
\rho_{\scriptscriptstyle
P}(\lambda|0_-)=\frac{\textrm{Tr}\Bigl[|\lambda\rangle\langle
\lambda|\hat{\rho}_{\scriptscriptstyle P}(0_-)\Bigr]}
{\textrm{Tr}\Bigl[\hat{\rho}_{\scriptscriptstyle P}(0_-)\Bigr]}=
\overline{\psi}^*(\lambda,0)\overline{\psi}(\lambda,0).
\end{equation}
If $\psi(\varphi,0)$ were given by the double Gaussian of (\ref{ijmpa.double}) then it is easy to prove
that also $\rho_{\scriptscriptstyle P}(\lambda|0_-)$ 
would be a double Gaussian with well defined mean values and finite 
standard deviations.
After the NSM of $\hat{\varphi}$ the probability distribution in $\lambda$ is given 
by:
\begin{equation}
\displaystyle \rho_{\scriptscriptstyle
M}(\lambda|0_+)=\frac{\textrm{Tr}\Bigl[|\lambda\rangle\langle\lambda|\hat{\rho}_{\scriptscriptstyle
M}(0_+)\Bigr]}{\textrm{Tr}\Bigl[\hat{\rho}_{\scriptscriptstyle M}(0_+)\Bigr]}
\end{equation}
which again is a uniform probability distribution; 
this is consistent with the fact that the superposition of 
the eigenstates $|\varphi_0\rangle$ of (\ref{mixin}) must correspond to a situation of total
ignorance for what concerns the variables $\lambda$ because of the commutation 
relation (\ref{comm}). So a measurement of $\hat{\varphi}$ in the
KvN theory
changes immediately the probability distribution in the conjugate variables $\lambda$, just like in QM 
a NSM of $\hat{q}$ changes immediately the probability distribution in the conjugate variables $p$. 

\smallskip

\item[{\bf b2)}] What we want to prove now is that, differently from what happens in QM, the probability
distributions in $\varphi$ do not change (with respect to the case {\bf a)} 
in which no measurement was done at $t=0$) not only at $t=0_+$ but also at any later time $t>0$. 
Performing the evolution of the statistical mixture (\ref{mixin}), we get at time $t=\tau$:
\begin{equation}
\displaystyle \hat{\rho}_{\scriptscriptstyle M}(\tau)=\int d\varphi_0|\psi(\varphi_0, 0)|^2
|\varphi_0,\tau\rangle\langle \varphi_0,\tau|.
\end{equation}
If the system we are considering is a free point particle then we have:
\begin{equation}
\displaystyle |\varphi_0,\tau\rangle=|q_0+p_0\tau/m,p_0\rangle.
\end{equation}
Once we represent the previous equation on $\langle\varphi|$ we obtain that in CM a Dirac delta in $\varphi$ remains 
a Dirac delta in $\varphi$ and it does not become a pure phase factor as in the quantum case (\ref{purephase}).
Consequently the probability density of finding $\varphi$ as result of a further measurement of
$\hat{\varphi}$ at time $t=\tau$ is given by:
\begin{equation}
\displaystyle \rho_{\scriptscriptstyle M}(\varphi|\tau)=
\frac{\textrm{Tr}\Bigl[|\varphi\rangle\langle\varphi|\hat{\rho}_{\scriptscriptstyle M}
(\tau)\Bigr]}{\textrm{Tr}\Bigl[\hat{\rho}_{\scriptscriptstyle M}(\tau)\Bigr]}=
|\psi(q-p\tau/m,p,0)|^2.\label{1633}
\end{equation}
The previous expression is just the modulus square of the evolution of the initial 
$\psi(\varphi,0)$ up to time 
$t=\tau$. This tells us that, differently than in QM, the probability density of the outcomes 
of $\hat{\varphi}$ is left unchanged by the NSM of $\hat{\varphi}$ 
not only at time $t=0_+$ but also at any later time $t>0$. 
In fact (\ref{1633}) is the same distribution as if the measurement at $t=0$ were not done.
This is true for every classical system and the reason  
is that in CM the equations of motion 
of $\varphi$ depend only on $\varphi$ and not on $\lambda$:
\begin{equation}
\displaystyle \dot{\varphi}^a=\omega^{ab}\frac{\partial H}{\partial \varphi^b}.
\end{equation}
Therefore at every time $t>0$ the
probability distribution in $\varphi$ 
will not feel $\lambda$ and, as a consequence, will not feel that the 
probability distribution in $\lambda$ was affected by the measurement at $t=0$. So 
we can say that the effect on $\lambda$ of the NSM of $\hat{\varphi}$ is not
inherited by $\varphi$ both at $t=0$ and at $t>0$. 
\end{itemize}

What we have done in the last two subsections can be summarized in the following scheme:

\bigskip
\bigskip

\begin{center}

\noindent {\Large{\bf Quantum Non Selective Measurement of $\hat{x}$:}}

\bigskip
\qquad {\Large{$\boxed{\;\textrm{Initial Quantum Wave Function}}\quad $}}\\

{\large $\begin{array}{cc}
 \big| \quad\quad & \quad\quad\quad \big| \\
\quad\quad \psi(x,0) \quad\quad   \quad\quad  &
\quad \qquad\textrm{NSM} \rightarrow\hat{\rho}_{\scriptscriptstyle M}(0) \\
 \Big| \quad\quad & \quad\quad\quad  \Big| \\
 \Big| \quad\quad & \quad\quad\quad  \Big| \smallskip \\
 \textrm{case {\bf a)}} \quad\quad & \quad\quad\quad  \textrm{case {\bf b2)}}\\
 \Big| \quad\quad & \quad\quad\quad  \Big| \\
 \Big\downarrow \quad\quad & \quad\quad\quad \Big\downarrow \smallskip\\
 \psi(x,\tau) \quad\quad & \quad\quad\quad \hat{\rho}_{\scriptscriptstyle M}(\tau)
\end{array}$}

 {\large \textrm{Outcome 1 for} $\hat{x} \quad\quad\;\;\; \neq$\quad\quad \textrm{Outcome 2 for} $\hat{x}$
}
\end{center}

\bigskip
\bigskip

\begin{center}

\noindent {\Large{\bf Classical Non Selective Measurement of $\hat{\varphi}$:}}

\bigskip
\qquad {\Large{$\boxed{\;\textrm{Initial Koopman-von Neumann Function}}\quad $}}\\

{\large $\begin{array}{cc}
 \big| \quad\quad & \quad\quad\quad \big| \\
\quad\quad \psi(\varphi,0) \quad\quad   \quad\quad  &
\quad \qquad\textrm{NSM} \rightarrow\hat{\rho}_{\scriptscriptstyle M}(0) \\
 \Big| \quad\quad & \quad\quad\quad  \Big| \\
 \Big| \quad\quad & \quad\quad\quad  \Big| \smallskip \\
 \textrm{case {\bf a)}} \quad\quad & \quad\quad\quad  \textrm{case {\bf b2)}}\\
 \Big| \quad\quad & \quad\quad\quad  \Big| \\
 \Big\downarrow \quad\quad & \quad\quad\quad \Big\downarrow \smallskip\\
 \psi(\varphi,\tau) \quad\quad & \quad\quad\quad \hat{\rho}_{\scriptscriptstyle M}(\tau)
\end{array}$}

 {\large \textrm{Outcome 1 for} $\hat{\varphi} \quad\quad\;\;\; =$\quad\quad \textrm{Outcome 2 for}
 $\hat{\varphi}$
}
\end{center}

\bigskip

We can conclude this section by saying that a NSM of $\hat{\varphi}$ in CM {\it does not disturb}
the probability distribution of $\varphi$ either immediately after the measurement or after a long time 
differently from what happens in QM for a NSM of $\hat{x}$. Therefore if we stick to the wisdom 
that the classical observables are only the Hermitian functions of $\hat{\varphi}$,
then we can say that a NSM of $\hat{\varphi}$ has no observable effect in CM.

\section{Two-slit experiment in CM and QM}

In \cite{waves1} we have already performed an analysis of the two-slit experiment 
showing, with a detailed calculation, how the interference figure is produced in QM and how, 
despite the fact that the KvN wave functions are complex, the interference figure cannot appear in the operatorial
formulation of CM. Nevertheless, in order to get a better understanding of the role of the evolution 
and of the crucial differences between classical and QM in the two-slit 
experiment, we want to further simplify the model analyzed in \cite{waves1}. In particular we want to strip 
off the calculation of all the mathematical details which are not necessary in explaining the
presence or not of an interference figure on the final screen. Quite surprisingly the
analysis that we want to propose is very similar to the one about non selective measurements 
analyzed in the previous section. 

First of all we want to briefly review the two-slit machinery analyzed in \cite{waves1}.
Let us consider a beam of particles propagating along the $y$ axis
and let us call $x$ the orthogonal axis. A first screen, with the two slits $\Delta_{\scriptscriptstyle 1}$
and $\Delta_{\scriptscriptstyle 2}$ centered respectively on $x_{\scriptscriptstyle A}=1$
and $-x_{\scriptscriptstyle A}=-1$, is labelled by the coordinate $y_{\scriptscriptstyle 0}$
along the $y$ axis. The final screen is labelled by the coordinate $y_{\scriptscriptstyle 1}$
like in the figure below:

\begin{center}
\begin{picture}(248,80)
\put(0,20){\makebox(0,0){$\scriptstyle{-x_A}$}}
\put(0,60){\makebox(0,0){$\scriptstyle{x_A}$}}
\put(16,80){\line(0,-1){80}}
\put(16,20){\makebox(0,0){$\bullet$}}
\put(16,60){\makebox(0,0){$\bullet$}}
\put(24,0){\makebox(0,0){$\scriptstyle{0}$}}
\put(230,80){\line(0,-1){80}}
\put(123,80){\line(0,-1){18}}
\put(120,62){\line(1,0){6}}
\put(120,58){\line(1,0){6}}
\put(123,58){\line(0,-1){36}}
\put(120,22){\line(1,0){6}}
\put(120,18){\line(1,0){6}}
\put(123,18){\line(0,-1){18}}
\put(135,20){\makebox(0,0){$\scriptstyle{\Delta_2}$}}
\put(135,60){\makebox(0,0){$\scriptstyle{\Delta_1}$}}
\put(142,0){\makebox(0,0){$\scriptstyle{y_0(t=0)}$}}
\put(249,0){\makebox(0,0){$\scriptstyle{y_1(t=1)}$}}
\end{picture}
\end{center}

\begin{figure}
\centering
\includegraphics[width=14cm]{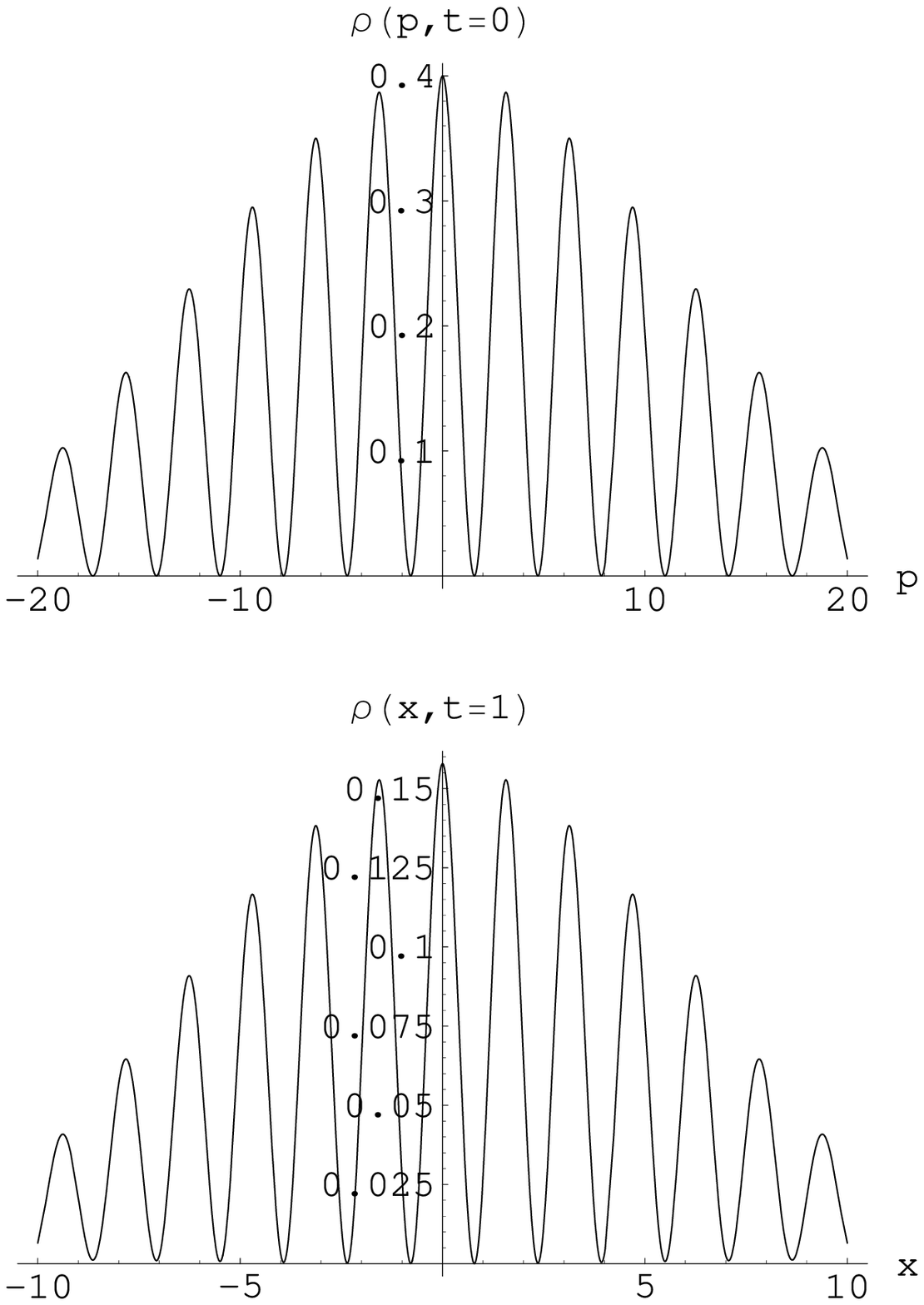}
\caption{\rm{The quantum two-slit experiment: the figure with maxima and minima
in $p$ at $t=0$ propagates at time $t=1$ to the distribution in $x$ creating the interference figure
on the final screen.}}
{\bf \label{simple}}
\end{figure}

Let us also suppose we know with arbitrary precision the position and the momentum 
of the particles along the $y$ direction at every instant of time $t$. This means that 
we know with arbitrary precision the time when the beam of particles arrives at the two
plates.
For example let us identify with $t=0$ the time when the beam of particles is near the plate 
with the two slits and with $t=1$ the time when the beam of particles reaches the final screen.
Let us then prepare a real initial quantum wave function reproducing the probability distribution 
of the particles near the two slits. For example let us assume a uniform distribution within the two slits:
\begin{equation}
\psi(x,t=0)=\left\{ 
\begin{array}{l}
\sqrt{\frac{5}{2}}\qquad -1.1\le x\le -0.9; \; 0.9\le x\le 1.1\medskip\\
0 \qquad\qquad \textrm{otherwise}
\end{array} \label{initialdis}
\right.  
\end{equation}
which implies
\begin{equation}
\rho(x,t=0)=\left\{
\begin{array}{l}
\frac{5}{2}\qquad -1.1\le x\le -0.9; \; 0.9\le x\le
1.1\medskip\\
0 \qquad\quad \textrm{otherwise}.
\end{array}
\right. 
\end{equation}
So at the beginning the distribution of particles along $x$ does not show any interference figure.
According to the rules of QM to obtain the probability distribution in the momentum space 
$p$ we have first to perform 
the Fourier transform of $\psi(x)$:
\begin{eqnarray}
\displaystyle \overline{\psi}(p,t=0)&\hspace{-0.2mm}=&\hspace{-0.2mm}
\frac{1}{\sqrt{2\pi\hbar}}\int_{-\infty}^{\infty}dx\,\psi(x,t=0) e^{-ipx/\hbar}=\nonumber\\
&\hspace{-0.2mm}=&\hspace{-0.2mm}\frac{1}{\sqrt{2\pi\hbar}}\int_{-1.1}^{-0.9}dx\,e^{-ipx/\hbar}
+\frac{1}{\sqrt{2\pi\hbar}}\int_{0.9}^{1.1}dx \,e^{-ipx/\hbar}. \label{gea}
\end{eqnarray}
In the RHS of (\ref{gea}) we have the same complex function, $\displaystyle e^{-ipx/\hbar}$, 
integrated over two different intervals.
This gives two complex functions with two different phases and, consequently, it produces 
a figure with maxima and minima in the
momenta distribution $|\overline{\psi}(p,t=0)|^2$. 
Now the equations of motion of $x$ depend explicitly on $p$ since
$\dot{x}=p/m$ and this implies that during the time evolution the figure with maxima and minima in $p$ is
inherited by the coordinates $x$ creating the well known interference effects. 
This emerges very clearly from {\bf Fig.}
{\bf \ref{simple}}, where we have put $\hbar=m=1$ and plot the probability distribution 
in $p$ at time $t=0$ and in $x$ at time $t=1$.

The situation is completely different in CM: in fact if
the initial KvN wave $\widetilde{\psi}(x,p)$, for the part in $x$, 
is given by (\ref{initialdis}), 
then the modulus square of its Fourier transform shows again a series of maxima and minima but 
now this modulus square is the probability density in $\lambda_x$. In fact, according
to the commutators (\ref{comm}), the variable canonically conjugated to $x$ is just 
$\lambda_x$. So it is the distribution 
$\rho(\lambda_x,t=0)$ which is identical to the $\rho(p,t=0)$ of {\bf Fig.}
{\bf \ref{simple}}.
The crucial difference is that in CM the variables $\lambda_x$ do not enter the equations of $x$. Therefore
the figure with maxima and minima in $\lambda_x$ is not inherited by $x$ during the motion
and there is no interference 
effect in $x$ on the final screen, like it happens in QM. There would be interference effects
in KvN theory if we could 
build a screen in $\lambda$-space or, equivalently, if we could admit, among the observables, 
the Hermitian operators depending also on $\lambda$. 

So we can say the following: the formal structure of KvN theory and QM is the same. Just after 
the first screen we do not have an interference figure in $x$ but a distribution well localized
near the two slits. Both in quantum and in CM the presence of the slits produces immediately
a figure with maxima and minima in the distribution of the variables conjugate to $x$ which are $p$ 
in QM 
and $\lambda_x$ in CM. Since only $p$ (and not $\lambda_x$) enters the equation
of motion of $x$ the figure with maxima and minima in the
conjugate variable is inherited by $x$ during the motion 
{\it only in the quantum case} creating the well known interference effects. This does not happen
in the classical case and as a consequence there is no interference at all on the final screen.
 
\section{Conclusions}

While in the first paper \cite{waves1} we looked at the role played by the phases of KvN ``wave functions"
in the different {\it representations} 
of the states, in this second paper
we sticked to the $\psi(\varphi)$ representation and showed that the postulate of KvN of having 
a Hilbert space of {\it complex} wave functions in CM implies the presence 
of ``generalized" observables depending not only on $\hat{\varphi}$ but also on $\hat{\lambda}$.
This is so if one wants that the phases associated with these wave functions be observable.
This makes the KvN theory a generalization of standard CM which could play a role 
in that region which lies at the interface between the classical and the quantum world.
The subtle differences between KvN theory and QM on the issue of measurements and two-slit
experiments have been also throughly studied in this paper.

\section*{Acknowledgments}

This work has been supported by grants from MIUR and INFN.

\end{document}